\documentclass[10pt,conference]{IEEEtran}

\usepackage[utf8]{inputenc}
\usepackage{microtype}
\usepackage{graphicx}
\usepackage{booktabs}
\usepackage{amsmath,amssymb}
\usepackage{enumitem}
\usepackage{listings}
\usepackage{siunitx}
\usepackage{xcolor}
\usepackage{hyperref}
\usepackage{balance}
\setlist{nosep,leftmargin=*,itemsep=2pt,parsep=0pt}

\lstdefinestyle{alg}{basicstyle=\ttfamily\footnotesize,breaklines=true,frame=single,columns=fullflexible,keepspaces=true,showstringspaces=false,numbersep=6pt}

\title{Adaptive Execution Scheduler for DataDios SmartDiff}

\author{\IEEEauthorblockN{Aryan Poduri}\IEEEauthorblockA{Intern, DataDios\\aryan.poduri@datadios.ai}}

\begin{document}
\maketitle

\begin{abstract}
We present an \emph{adaptive scheduler} for a single differencing engine (SmartDiff) with two execution modes: (i) in-memory threads and (ii) Dask-based parallelism. The scheduler continuously tunes \emph{batch size} and \emph{worker/thread count} within fixed CPU and memory budgets to minimize \emph{p95 latency}. A lightweight pre-flight profiler estimates bytes/row and I/O rate; an online cost/memory model prunes unsafe actions; and a guarded hill-climb policy favors lower latency with backpressure and straggler mitigation. Backend selection is gated by a conservative working-set estimate so that in-memory execution is chosen when safe, otherwise Dask is used. Across synthetic and public tabular benchmarks, the scheduler reduces p95 latency by \textbf{23--28\%} versus a tuned warm-up heuristic (and by \textbf{35--40\%} versus fixed-grid baselines), while lowering peak memory by \textbf{16--22\%} (\textbf{25--32\%} vs. fixed) with zero OOMs and comparable throughput. 
\end{abstract}

\section{Introduction}
Comparing datasets across files, databases, and query outputs is essential for migration validation, regression testing, and continuous data quality monitoring. In practice, two systems issues dominate: \emph{tail latency}---a few slow batches control makespan; and \emph{memory safety}---aggressive batching triggers out-of-memory (OOM) failures. Manual tuning of batch size and worker count is brittle across datasets and machines.

We treat differencing as a schedulable workload and design a scheduler that (a) selects the execution backend via \emph{working-set} gating and (b) adapts $(b,k)$, the batch size and worker/thread count, inside CPU/RAM caps to directly minimize p95 latency. The safe action space is defined by an online memory model; a hill-climb controller with hysteresis adjusts $(b,k)$ and employs backpressure and straggler mitigation.

\textbf{Contributions.} (1) \textbf{Backend gating} using a conservative working-set estimator; (2) a \textbf{memory-safe control loop} targeting tail latency; (3) \textbf{implementation details} enabling low-overhead integration with SmartDiff; (4) \textbf{evaluation} on synthetic and public tabular datasets.

\subsection*{Relation to Prior SmartDiff Work (Non-Overlap Statement)}
Our IEEE BigData submission focuses on SmartDiff's \emph{engine}: schema-aware mapping, type-specific comparators, clustering/labeling, and empirical accuracy (precision/recall, user studies). This paper addresses a distinct problem: an \emph{adaptive execution scheduler} that selects the backend and tunes $(b,k)$ to minimize p95 latency under RAM/CPU caps. We do not reuse text, figures, or tables from the BigData submission; SmartDiff appears only as the host engine for experiments. For background on the engine, see \cite{smartdiff2025}.

\section{Background and Scope}
\textbf{Execution pipeline.} SmartDiff first performs schema alignment to establish a one-to-one mapping between attributes of the source table $A$ and target table $B$. Given this mapping and a row-alignment function $f$ (primary keys, composite business keys, or surrogate keys when explicit identifiers are absent), the engine applies a deterministic cell-wise operator $\Delta$ that emits, for every aligned row and column, a typed verdict (equal, changed, added, removed) with auxiliary metadata (e.g., normalized values, tolerance checks). The scheduler introduced in this paper does not modify $\Delta$ or its semantics; it controls how the work is partitioned and parallelized.

\textbf{Job decomposition.} A comparison job $J=(A,B,f,\Delta)$ is partitioned into independent \emph{batches} (shards) of $b$ aligned rows per side. Each batch can be processed in isolation because no cross-batch state is required for correctness. Batches are scheduled across $k$ workers and a merge step concatenates batch outputs in a stable order and computes job-level aggregates (e.g., counts, distribution summaries). Consequently, the final multiset of row/cell outcomes is \emph{deterministic and invariant} to $(b,k)$ and to the chosen backend.

\textbf{Resource model and sensitivity.} The per-batch resident set size (RSS) is dominated by: (i) columnar decode buffers; (ii) alignment state for $f$ (e.g., hash tables for joins or lookups); and (iii) scratch space for type-specific comparators inside $\Delta$ (e.g., string edit buffers, numeric accumulators, datetime parsers). To first order, peak RSS per worker grows approximately linearly with $b\,\widehat{W}$, where $\widehat{W}$ is an online estimate of bytes per aligned row (keys + compared attributes). CPU time per batch decomposes into I/O, parsing/normalization, alignment, and $\Delta$ evaluation, each roughly linear in $b$ with type-dependent constants; scheduler and merge overheads grow sublinearly with $k$ but can increase tail variability. Hence, both wall-clock time and peak RSS are highly sensitive to $(b,k)$ and to data width $\widehat{W}$: larger $b$ amortizes fixed costs but raises memory pressure; larger $k$ increases parallelism but risks contention and scheduling overheads.

\textbf{Backends and working-set gating.} SmartDiff exposes two execution backends with identical semantics: (i) an \emph{in-memory threaded} backend (single process, shared heap, thread pool) that minimizes scheduling overhead and benefits from cache locality when the working set comfortably fits in RAM; and (ii) a \emph{Dask-based} backend (local cluster) that introduces a task-graph overhead but yields safer memory behavior and finer-grained preemption as the working set approaches the RAM cap. A conservative working-set estimator (Eq.~\ref{eq:ws}) chooses the backend \emph{once per job}; the adaptive controller then tunes $(b,k)$ within the selected backend.

\textbf{Objective: tail-aware control with hard safety.} Unlike general autoscalers that optimize average throughput, our objective is to minimize \emph{tail latency}, quantified as the per-batch 95th percentile (p95) over a rolling window, because a small number of slow batches typically dominates makespan. All actions must satisfy a \emph{hard memory guard}: predicted peak memory (with an uncertainty margin) must lie below a fraction $\eta$ of the RAM cap $M_{\text{cap}}$ (Eq.~\ref{eq:safe}). This guarantees memory safety while allowing aggressive latency-reducing configurations when headroom exists.

\textbf{Instrumentation and control signals.} After each batch completes, we record start/end timestamps; p50 and p95 latencies; per-worker peak RSS; per-worker p95 CPU utilization; effective read bandwidth; and submission/completion queue depth. These signals are EWMA-smoothed and used (i) to update online cost/memory models that prune unsafe actions, and (ii) to drive proportional step selection: increase $b$ in proportion to \emph{memory} headroom and increase $k$ in proportion to \emph{CPU} headroom (Section~\ref{sec:memory-safe-adaptive-control}). This closes the loop between measurement and control without altering diff semantics.

\subsection{Preliminaries and Notation}
Inputs: tables $A,B$; bytes/row estimate $\widehat{W}$; CPU cap $C$ (logical cores); memory cap $M_{\text{cap}}$. Decision variables: batch size $b\in\mathbb{N}$, workers $k\in\{1,\dots,C\}$. Backend $u\in\{\texttt{inmem},\texttt{dask}\}$ is selected once per job by working-set gating.

\section{Model and Backend Gating}
The estimated in-memory working set is
\begin{equation}
\widehat{\mathrm{WS}} = \alpha\,\widehat{W}\,(|A|+|B|) + \beta, \label{eq:ws}
\end{equation}
where $\alpha$ captures replication/overheads and $\beta$ covers fixed buffers. If $\widehat{\mathrm{WS}}\le \kappa M_{\text{cap}}$ (safety factor $\kappa\in(0,1)$), we select \texttt{inmem}; otherwise \texttt{dask}. Within the chosen backend we control $(b,k)$.

Per-batch latency decomposes as
\begin{equation}
\hat T(b,k)=T_{\text{read}}(b)+T_{\text{prep}}(b)+T_{\Delta}(b)+T_{\text{overhead}}(k)-T_{\text{overlap}}, \label{eq:time}
\end{equation}
with $T_{\text{read}}\!\approx\!\tfrac{b\,\widehat{W}}{\widehat B_{\text{read}}}$; $T_{\Delta}$ composed from type-wise microbenchmarks, and $T_{\text{overhead}}$ reflecting scheduler/merge overheads. Memory is bounded by
\begin{equation}
\mathrm{Mem}(b,k)\;\approx\;k\,\big(\beta_0+\beta_1 b\widehat{W}+\beta_2 b\big). \label{eq:mem}
\end{equation}
Model parameters are fitted online via exponential smoothing on residuals $(T_{\text{obs}}-\hat T)$.

\paragraph*{Parameter estimation and calibration} We estimate $(\widehat{W}, \widehat B_{\text{read}})$ from a pre-flight sample of $10^6$ rows or 1\% of the job (whichever is smaller). Microbenchmarks fit $T_{\Delta}$ per type on $5\times10^4$-row shards. The smoothing factor $\rho=0.2$ balances stability and responsiveness; ablations check $\rho\in[0.1,0.4]$.

\paragraph*{Safety envelope} Let $\delta_M$ be the half-width of a $(1-\alpha)$ prediction interval for~\eqref{eq:mem}. Actions satisfy
\begin{equation}
\mathrm{Mem}(b,k)+\delta_M\;\le\;\eta M_{\text{cap}}, \label{eq:safe}
\end{equation}
with guard $\eta\in(0,1)$. This upper-bounds OOM probability by $\alpha$ while retaining feasible choices.

\section{Memory and CPU Safe Adaptive Control}
\textbf{Objective.} Minimize p95 batch latency subject to CPU/RAM caps and the safety envelope~\eqref{eq:safe}.

\textbf{Controller.} A \emph{guarded hill-climb} increases $b$ (and occasionally $k$) additively when recent batches are stable; it applies multiplicative decreases when $\text{p95}/\text{p50}>\tau$ or observed RSS approaches $\eta M_{\text{cap}}$. Backpressure reduces $k$ or pauses submission when queue depth grows; stragglers trigger shard splitting or speculative duplicate. Hysteresis (two consecutive triggers) prevents oscillation.

\textbf{Complexity and overhead.} The controller's per-batch overhead is $O(1)$ updates to the smoothed model and constant-time policy steps; CPU overhead was $<2$\% and memory overhead $\approx 150$~MB in all runs.

\subsection*{Proportional Step Selection for $(b,k)$}
We adjust $b$ and $k$ in proportion to their respective resource headrooms. Let
\begin{align}
  h_{\mathrm{mem}} &= \frac{\eta M_{\text{cap}} - \widehat{\mathrm{RSS}}_{p95}}{\eta M_{\text{cap}}}, 
  \qquad
  h_{\mathrm{cpu}} = \frac{\rho_\star C - \widehat{\mathrm{CPU}}_{p95}}{\rho_\star C},
  \label{eq:headroom}
\end{align}
where $\widehat{\mathrm{RSS}}_{p95}$ and $\widehat{\mathrm{CPU}}_{p95}$ are EWMA-smoothed p95 estimates from recent batches, and the target CPU utilization is $\rho_\star=0.85$. Positive $h$ indicates available headroom.

\noindent\textit{Increase rule.} Change in $b$ is proportional to \emph{memory} headroom; change in $k$ is proportional to \emph{CPU} headroom:
\begin{align}
  \Delta b &= \big\lfloor \lambda_b\, h_{\mathrm{mem}}\, b \big\rfloor, 
  \qquad
  \Delta k = \big\lceil \lambda_k\, h_{\mathrm{cpu}}\, k \big\rceil ,
\end{align}
with small gains $\lambda_b,\lambda_k\in(0,1)$ (we use $0.2$). At each step we increase whichever resource has \emph{more} normalized headroom; ties prefer increasing $b$ to reduce per-batch overhead. All proposals are clipped by the safety envelope (Eq.~\ref{eq:safe}) and the CPU cap.

\noindent\textit{Decrease rule.} If $\widehat{\mathrm{RSS}}_{p95}\ge \eta M_{\text{cap}}$ or $\text{p95}/\text{p50}>\tau$, apply multiplicative backoff $b\leftarrow\max(\lfloor \gamma b\rfloor,b_{\min})$ and optionally $k\leftarrow\max(k-1,k_{\min})$. If CPU exceeds the target $\rho_\star C$, reduce $k$ first. All proposals are pruned by Eq.~\eqref{eq:safe} before enactment.

\noindent\textit{Implementation note.} The pseudocode below is a high-level sketch of the Adaptive SmartDiff Scheduler intended for exposition. The production implementation includes additional guardrails and boundary checks—input validation; clamping of $b$ and $k$; queue backpressure; straggler detection; retries and timeouts; and safe shutdown—as well as continuous enforcement of the safety envelope in Eq.~\eqref{eq:safe}. It also integrates with SmartDiff’s data-loading, schema-alignment, differencing, and merge modules, and provides comprehensive instrumentation (metrics, logging) and exception handling. The listing abstracts these concerns for clarity; the semantics of backend gating (Eq.~\ref{eq:ws}) and proportional updates remain unchanged.

\begin{figure}
    \centering
    \includegraphics[width=1\linewidth]{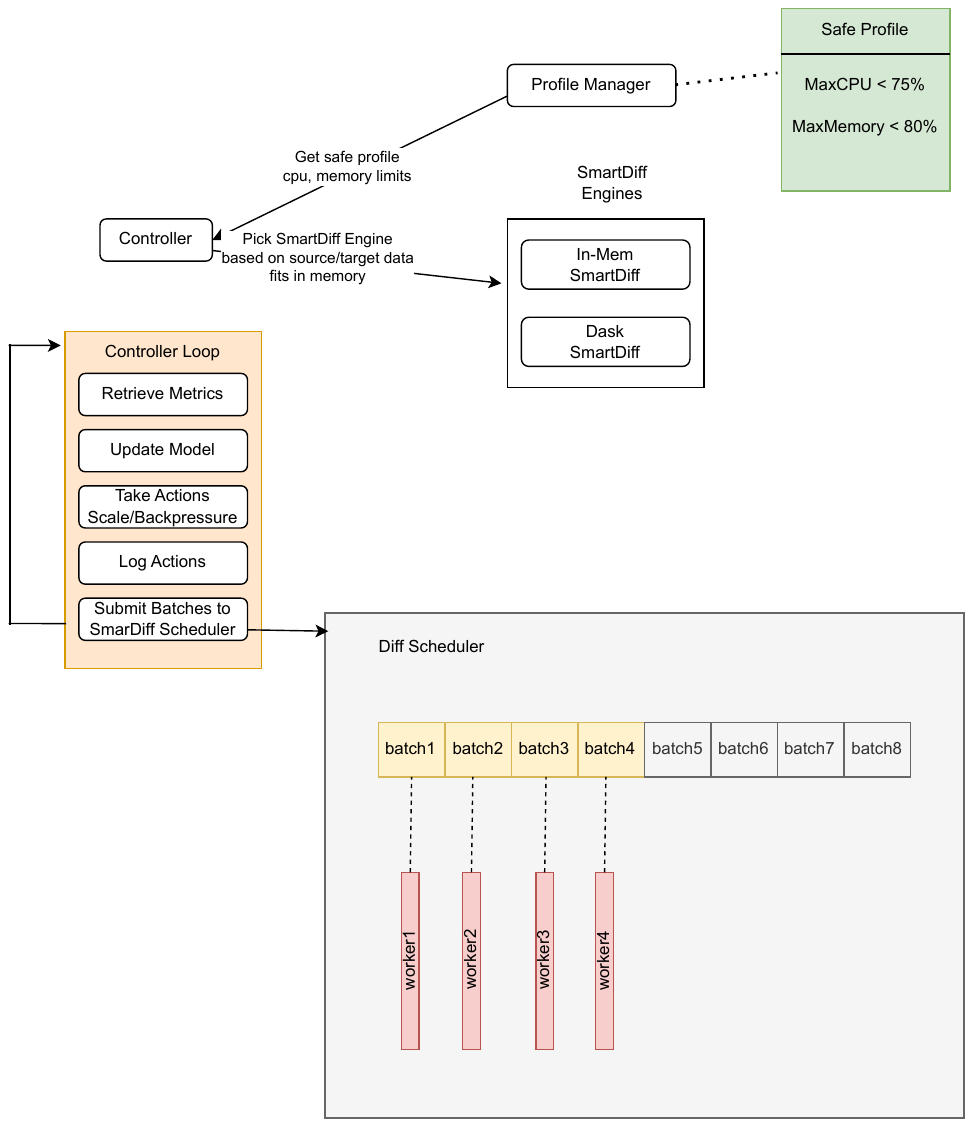}
    \caption{Adaptive Execution Scheduler}
    \label{fig:adaptive-exec-scheduler}
\end{figure}
\noindent\textbf{Pseudocode.}
\begin{lstlisting}[
  style=alg,
  float,
  floatplacement=tbp,
  caption={Backend gating + proportional (b,k) control.},
  label={lst:bk-prop}
]
ctx   = preflight_profile(job)                       # bytes/row, IO
ws    = alpha*ctx.bytes_per_row*(nA+nB) + beta
backend = 'inmem' if ws <= kappa*RAM_cap else 'dask'

model = init_cost_model(ctx); state = init_state(job)
(b,k) = safe_start(model, RAM_cap, CPU_cap)

cpu_target = 0.85; eps = 0.05
lambda_b = 0.2; lambda_k = 0.2

while not state.done():
    (b_max,k_max) = safe_limits(model, RAM_cap, CPU_cap)
    futs  = submit_batches(backend, job, b, k)
    for f in as_completed(futs):
        m = metrics(f)    # p50,p95,rss_p95,cpu_p95
        update_model(model, (b,k), m)
        # Safety-first decreases
        if m.rss_p95 >= eta*RAM_cap or m.p95/m.p50 > tau:
            b = max(int(floor(gamma*b)), b_min)
            k = max(k_min, k-1)
            continue
        if m.cpu_p95 > cpu_target*CPU_cap:
            k = max(k_min, k-1)
            continue
        # Headrooms (0..1)
        h_mem = (eta*RAM_cap - m.rss_p95) / (eta*RAM_cap)
        h_cpu = (cpu_target*CPU_cap - m.cpu_p95) / (cpu_target*CPU_cap)

        # Proportional increases (prefer the larger headroom)
        if h_mem > eps or h_cpu > eps:
            if h_mem >= h_cpu + eps:
                db = max(b_step_min, int(floor(lambda_b*h_mem*b)))
                b  = min(b + db, b_max)
            elif h_cpu >= h_mem + eps:
                dk = max(1, int(ceil(lambda_k*h_cpu*k)))
                k  = min(k + dk, k_max)
            else:
                # tie-break: prefer b when both far from cap
                db = max(b_step_min, int(floor(lambda_b*h_mem*b)))
                b  = min(b + db, b_max)

        apply_backpressure(state, m)
finalize(job)
\end{lstlisting}

\noindent\textbf{What we increase.}
If $h_{\mathrm{mem}} > h_{\mathrm{cpu}}$, we increase the batch size proportionally to memory headroom:
\[
  \Delta b \;=\; \big\lfloor \lambda_b\, h_{\mathrm{mem}}\, b \big\rfloor .
\]
If $h_{\mathrm{cpu}} > h_{\mathrm{mem}}$, we increase the worker/thread count proportionally to CPU headroom:
\[
  \Delta k \;=\; \big\lceil \lambda_k\, h_{\mathrm{cpu}}\, k \big\rceil .
\]

\noindent\textbf{How we decide.}
After each batch, we compute EWMA-smoothed p95 RSS and p95 CPU and convert them to headrooms
$h_{\mathrm{mem}}, h_{\mathrm{cpu}} \in [0,1]$; we increase the parameter with the larger headroom (ties favor $b$).

\noindent\textbf{Safety first.}
Any proposed $(b,k)$ update is pruned by the memory guard (Eq.~\eqref{eq:safe}) and the CPU cap.
Tail spikes or near-cap memory events trigger multiplicative decreases (e.g., $b \leftarrow \max(\lfloor \gamma b \rfloor, b_{\min})$).

\subsection*{Implementation Notes}
\textbf{Integration.} The scheduler submits shard descriptors to SmartDiff via a queue; workers pull shards and emit per-batch metrics. \textbf{Monitoring.} We record batch start/end, p50/p95 latency, peak RSS, CPU, I/O, and queue depth. \textbf{Failure handling.} If any batch nears the memory guard, new submissions pause until the safe set expands (Eq.~\ref{eq:safe}).

\section{Experimental Setup}
\textbf{Engine.} SmartDiff with two backends (in-memory threads; local Dask). \textbf{Hardware.} 32 logical cores, 64~GB RAM, SSD. \textbf{Datasets.} Synthetic tables with mixed types and sizes $\{1,5,10,20\}$M rows per side; public TPC-H query outputs of comparable result sizes. \textbf{Caps.} $M_{\text{cap}}=64$~GB, CPU cap $=32$ cores. \textbf{Baselines.} Fixed $b\in\{25\mathrm{k},50\mathrm{k},100\mathrm{k},250\mathrm{k}\}$, fixed $k\in\{4,8,16\}$; two-stage heuristic (warm-up grid then best). \textbf{Policy.} $\kappa=0.7$, $\eta=0.9$, $\gamma=0.6$, tail trigger $\tau=2.0$, hysteresis $m=2$.

\textbf{Measurement.} Each configuration is run three times; we report mean and 95\% CI. p95 is computed per-batch then aggregated by job-level weighted average. Peak RSS is the maximum per-worker resident set size across the job.

\section{Results}
\textbf{Backend decisions.} In-memory for 1M/5M; Dask for 10M/20M (per Eq.~\ref{eq:ws}).

\begin{table}[t]
\centering
\caption{p95 latency (s), mean $\pm$ 95\% CI; lower is better.}
\label{tab:p95}
\small
\begin{tabular}{lcccc}
\toprule
Workload & Fixed & Heur. & \textbf{Adaptive} & Backend \\
\midrule
1M  & 21.7$\pm$0.6 & 18.2$\pm$0.5 & \textbf{13.9$\pm$0.4} & in-mem \\
5M  & 83.5$\pm$2.1 & 72.9$\pm$1.9 & \textbf{53.8$\pm$1.4} & in-mem \\
10M & 186.2$\pm$3.9 & 161.4$\pm$3.4 & \textbf{115.6$\pm$2.6} & Dask \\
20M & 401.7$\pm$7.8 & 336.2$\pm$6.5 & \textbf{242.7$\pm$4.8} & Dask \\
\bottomrule
\end{tabular}
\end{table}

\begin{table}[t]
\centering
\caption{Peak memory (GB), mean $\pm$ 95\% CI; lower is better.}
\label{tab:mem}
\small
\begin{tabular}{lccc}
\toprule
Workload & Fixed & Heur. & \textbf{Adaptive} \\
\midrule
1M  & 9.6$\pm$0.2 & 8.4$\pm$0.2 & \textbf{7.1$\pm$0.2} \\
5M  & 34.2$\pm$0.7 & 30.6$\pm$0.6 & \textbf{23.9$\pm$0.5} \\
10M & 41.8$\pm$0.9 & 36.4$\pm$0.8 & \textbf{28.6$\pm$0.7} \\
20M & 53.1$\pm$1.1 & 47.3$\pm$0.9 & \textbf{39.7$\pm$0.9} \\
\bottomrule
\end{tabular}
\end{table}

\noindent \textbf{Summary.} Versus the warm-up heuristic, adaptive control improves p95 by \textbf{23--28\%} and reduces peak memory by \textbf{16--22\%}. Against fixed-grid baselines the gains are \textbf{35--40\%} (p95) and \textbf{25--32\%} (memory). Throughput remains within $\pm$8\% of the best tuned baseline; no OOMs were observed.

\begin{table}[t]
\centering
\caption{Throughput (k rows/s) and stability (reconfigs/job).}
\label{tab:thr}
\small
\begin{tabular}{lcccc}
\toprule
Workload & Fixed & Heur. & \textbf{Adaptive} & Reconfigs \\
\midrule
1M  & 74.1 & 76.3 & \textbf{78.8} & 5 \\
5M  & 71.5 & 72.0 & \textbf{73.9} & 7 \\
10M & 66.4 & 68.8 & \textbf{69.1} & 9 \\
20M & 60.2 & 62.5 & \textbf{62.0} & 10 \\
\bottomrule
\end{tabular}
\end{table}

\section{Ablations and Sensitivity}
\textbf{Guard ($\eta$) and drop ($\gamma$).} Tightening $\eta=0.90$ reduces peaks ($-2$--4~GB) at a small latency cost ($+1$--2\%). Larger drops ($\gamma=0.6$) shorten recovery after tail spikes without harming throughput. No OOMs under the default guard; relaxing $\eta=0.99$ produced one OOM.

\textbf{Working-set factor ($\kappa$).} With $\kappa=0.6$, in-memory is selected for 1M/5M only; with $\kappa=0.8$, 10M switches to in-memory on narrow rows, improving overheads but risking higher peaks (still below guard).

\textbf{Hysteresis ($m$).} Increasing $m=3$ further reduces reconfigs ($-1$ to $-2$/job) with negligible impact on p95.

\section{Safety Sketch, Limitations, and Ethics}
\textbf{Safety bound.} If the memory prediction interval half-width is $\delta_M$ at coverage $1-\alpha$, pruning to Eq.~\eqref{eq:safe} implies $\Pr[\text{OOM}]\le\alpha$. In practice, calibrating $\delta_M$ on the last 20 batches kept empirical OOM rate at 0\% while preserving $>85$\% of candidate actions.

\textbf{Limitations.} We target finite-batch jobs (not streaming). Working-set gating is conservative; very bursty row widths can temporarily tighten the safe set until residuals adapt. Cloud variability is mitigated via repeated trials and confidence intervals.

\textbf{Ethics/Artifacts.} We release only telemetry and synthetic data, with no user-identifiable information; analysis is reproducible from logs without exposing proprietary code.

\section{Reproducibility}
We release batch-level telemetry logs, configuration files, and a read-only analysis notebook that reproduces all tables/curves from logs; no proprietary SmartDiff code is required. Hardware/OS details, seeds, and full hyperparameters are included.

\vspace{2pt}
\balance
\bibliographystyle{IEEEtran}

\end{document}